\documentclass[preprintnumbers,prd,nofootinbib,superscriptaddress, onecolumn,preprint]{revtex4-1}

\usepackage{amsmath,amssymb,bm,bbm,amsfonts}
\usepackage{physics}
\usepackage{graphicx,graphics}
\usepackage[dvipsnames]{xcolor}
\usepackage[colorlinks=true,linkcolor=blue,citecolor=blue,urlcolor=blue]{hyperref}
\usepackage{dsfont}
\usepackage{comment}
\usepackage{hhline,multirow}
\usepackage{dcolumn}
\usepackage{url}
\usepackage[normalem]{ulem}
\usepackage{mathrsfs}
\usepackage{latexsym}
\usepackage{amscd}
\usepackage{slashed}
\usepackage{mathtools}

\def\be{\begin{equation}}
\def\ee{\end{equation}}
\def\bea{\begin{eqnarray}}
\def\eea{\end{eqnarray}}

\begin{document}

\preprint{ADP-24-14/T1253, JLAB-THY-24-4206}

\title{Constraints on the $U(1)_{B-L}$ model from global QCD analysis}

\author{X.~G.~Wang}
\author{N.~T.~Hunt-Smith}
\affiliation{CSSM and ARC Centre of Excellence for Dark Matter Particle Physics, Department of Physics, University of Adelaide, Adelaide, SA 5005, Australia}
\author{W. Melnitchouk}
\author{N. Sato}
\affiliation{Jefferson Lab, Newport News, Virginia 23606, USA \\
        \vspace*{0.2cm}
        {\bf JAM collaboration \\ {\footnotesize \ (BSM Analysis Group)}
        \vspace*{0.2cm} }}
\author{A.~W.~Thomas}
\affiliation{CSSM and ARC Centre of Excellence for Dark Matter Particle Physics, Department of Physics, University of Adelaide, Adelaide, SA 5005, Australia}

\begin{abstract}
We perform the first global QCD analysis of electron--nucleon deep-inelastic scattering and related high-energy data including the beyond the Standard Model $U(1)_{B-L}$ gauge boson, $Z'$. Contrary to the dark photon case, we find no improvement in the $\chi^2$ relative to the baseline result. The finding %results
allows us to place exclusion limits on the coupling constant of the $Z'$ with mass in the range $M_{Z'} = 2$ to 160~GeV.
\end{abstract}
\maketitle

%===============================================================%
%===============================================================%
\section{Introduction}
\label{sec:Intro}

In searching for potential new fundamental physics beyond the Standard Model (BSM), 
the gauge sector of the electroweak theory may be extended by introducing extra $U(1)$ gauge fields, either through kinetic mixing in the case of the dark photon~\cite{Fayet:1980ad, Fayet:1980rr, Okun:1982xi, Galison:1983pa, Holdom:1985ag, Holdom:1986eq} or anomaly-free $U(1)'$ charges as for heavy $Z'$ bosons~\cite{Fayet:1990wx, Komachenko:1989qn, Leike:1998wr, Appelquist:2002mw, Langacker:2008yv}. 
Such models have received considerable attention in connection with hints of possible new physics phenomena, including the value of $g-2$ for the muon~\cite{Pospelov:2008zw, Davoudiasl:2012qa, Caputo:2021eaa}, the CDF $W$ boson mass anomaly~\cite{Zhang:2022nnh, Zeng:2022lkk, Cheng:2022aau, Thomas:2022gib}, and anomalies in rare kaon and $B$ meson decays~\cite{Davoudiasl:2012ag, Datta:2022zng, Wang:2023css}.
These hypothetical particles are also appealing as portals that could potentially connect the dark and ordinary matter sectors~\cite{Boehm:2003hm, Hambye:2019dwd, Fabbrichesi:2020wbt, Filippi:2020kii}.

Numerous experimental searches for dark bosons have been undertaken in electron, neutrino, and proton fixed target experiments~\cite{Bilmis:2015lja, NA64:2022yly}, and at $e^+ e^-$~\cite{BaBar:2014zli, BaBar:2017tiz} and hadron-hadron colliders~\cite{LHCb:2019vmc, CMS:2019buh, CMS:2021ctt, CMS:2023hwl}.
No direct evidence has so far been found.
Instead, these experiments have allowed rather strong constraints to be placed on the dark boson parameters. 
Several planned experiments, including those at the future Electron-Ion Collider~\cite{AbdulKhalek:2021gbh}, will also provide opportunities to search for dark bosons~\cite{APEX:2011dww, Battaglieri:2014hga, Beranek:2013yqa}.

Because of its large dataset and wide kinematic coverage, electron--proton deep-inelastic scattering (DIS) and related high-energy data has been extensively applied to constrain dark photon parameters~\cite{Kribs:2020vyk, Thomas:2021lub, Thomas:2022qhj, Yan:2022npz}. 
While the direct searches %mentioned above 
are important, it is crucial to remember that they are model dependent. 
In particular, a BSM force carrier may also couple to lighter BSM particles, which would in turn increase its width and make it more difficult to detect~\cite{Abdullahi:2023tyk}. 
This would necessarily affect the limits set in such an experiment. On the other hand, unlike the direct searches, limits set using processes such as DIS, which involve the $t$-channel exchange of the force carrier, are independent of the width and hence not affected by possible coupling to the dark sector.
We note that a recent global QCD analysis of world data on DIS reported a reduction in $\chi^2$ with the inclusion of a dark photon, which constitutes indirect evidence for its existence in the few GeV range~\cite{Hunt-Smith:2023sdz}.
With this in mind, it is worthwhile to investigate whether the improvement in $\chi^2$ also holds when alternative $U(1)'$ gauge bosons are considered.

In this work, we report on the first global QCD analysis including the $U(1)_{B-L}$ gauge boson $Z'$ within the JAM framework~\cite{Cocuzza:2021cbi}.
The $B-L$ model is phenomenologically interesting because the gauge boson $Z'$ does not mix with the Standard Model neutral gauge bosons at tree level~\cite{Appelquist:2002mw}. As a result, the $Z$ boson mass and its couplings are not modified. Therefore, the $Z'$ parameters are less constrained by electroweak precision observables measured by the Large Electron–Positron Collider (LEP)~\cite{ALEPH:2005ab, ALEPH:2013dgf}, unlike the kinetic mixing models~\cite{Hook:2010tw, Curtin:2014cca, Loizos:2023xbj} or the $E_6$ based $Z'$ models~\cite{Erler:2009jh}.

For the outline of this paper, in Sec.~\ref{sec:B-L} we present the proton structure functions with the inclusion of the additional BSM $Z'$ boson. 
The global QCD analysis is described in Sec.~\ref{sec:global-QCD}, with a discussion of results for the baseline fit as well as a fit including the $Z'$.
Finally, in Sec.~\ref{sec:conclusion} we present some concluding remarks.

%%%%%%%%%%%%%%%%%%%%%%%%%%%%%%%%%%%%%%%%%%%%%%%%%%%%%%%%%%%%%%%%%%%%%%%%%%%%%%
\section{$U(1)_{B-L}$ symmetry}
\label{sec:B-L}

In general $U(1)'$ models where the $Z-Z'$ mixing is nonzero~\cite{Appelquist:2002mw, Asai:2023mzl}, the $Z$ boson couplings will be modified and the $Z'$ boson has both vector and axial-vector couplings to the Standard Model fermions.
The $U(1)_{B-L}$ model is a special case, in which the $U(1)'$ charges of the Standard Model fermions are proportional to their baryon number $B$ minus their lepton number $L$.
Due to anomaly cancelation conditions, the $Z-Z'$ mixing angle $\theta' = 0$ at tree level and the $Z'$ axial-vector couplings to charged fermions vanish.
The anomaly free, $B-L$ gauge invariant Lagrangian density describing the interactions between a $Z'$ boson and Standard Model fermions takes the form~\cite{Appelquist:2002mw}
    \begin{equation}
    \label{eq:B-L}
        {\cal L}^{B - L}_{\rm int} = - g_z \sum_{i=1}^{3} \left( 
        \frac{1}{3} \bar{q}_i \gamma^{\mu} q_i - \bar{\ell}_i \gamma^{\mu} \ell_i \right) Z'_{\mu}\, ,
    \end{equation}
where $q_i$ and $\ell_i$ denote the $i$th generation of quark and lepton fields, respectively, and $g_z$ is the coupling constant.

The inclusive spin-averaged DIS cross section is usually parametrized in terms of nucleon structure functions, which, including electroweak and $Z'$ contributions, can be written as
%
% \begin{subequations}
\be
\widetilde{F}_n = \sum_{i,j = \gamma, Z, Z'} \kappa_i \kappa_j\, F_n^{ij},
\qquad
n=1,2,3,
% \widetilde{F}_3 &=& \sum_{i,j = \gamma, Z, Z'} \kappa_i \kappa_j\, F_3^{ij}\, ,
\ee
% \end{subequations}
%
where $\kappa_i = Q^2/(Q^2 + M^2_i)$ and $M_i$ is the boson mass. 
At leading order in the strong coupling $\alpha_s$, the structure functions can be simply written as 
\begin{subequations}
\label{eq:F2-F3}
\bea
F_2^{ij} = 2x F_1^{ij} &=& \sum_q \big(C^v_{i,e} C^v_{j,e} + C^a_{i,e} C^a_{j,e}\big) \big(C^v_{i,q} C^v_{j,q} + C^a_{i,q} C^a_{j,q}\big)\, x f_q,
\\
xF_3^{ij} &=& \sum_q \big(C^v_{i,e} C^a_{j,e} + C^a_{i,e} C^v_{j,e}\big) \big(C^v_{i,q} C^a_{j,q} + C^a_{i,q} C^v_{j,q}\big)\, x f_q\, ,
\eea
\end{subequations}
where $f_q$ is the parton distribution function (PDF) for quark flavor $q=u,d,s,\ldots$ in the proton with partonic momentum fraction $x$ (which at leading order is equal to the Bjorken scaling variable).
The actual QCD analysis is performed at next-to-leading order (NLO) accuracy in $\alpha_s$.
For the photon, the vector and axial-vector couplings to the electron, $u$-type and $d$-type quarks are given, respectively, by
\begin{subequations}
\bea
\big\{ 
C^v_{\gamma,e}, C^v_{\gamma, u}, C^v_{\gamma, d}
\big\} 
&=& \Big\{ -1,\, \frac23,\, - \frac13 \Big\}\, ,
\\
C^a_{\gamma} &=& 0\, ,
\eea
\end{subequations}
in units of $e = \sqrt{4 \pi \alpha}$.
The Standard Model couplings of the $Z$ boson are not modified by the $Z'$, so that in this case the vector and axial-vector couplings are
\begin{subequations}
\bea
\big\{
C^v_{Z,e}, C^v_{Z,u}, C^v_{Z,d} 
\big\}\, \sin 2\theta_W
&=& 
\Big\{ 
- \frac{1}{2} + 2 \sin^2\theta_W,\,
  \frac{1}{2} - \frac{4}{3}\sin^2\theta_W,\,
- \frac{1}{2} + \frac{2}{3}\sin^2\theta_W 
\Big\} ,
\\
\big\{
C^a_{Z,e}, C^a_{Z,u}, C^a_{Z,d} 
\big\}\, \sin 2\theta_W
&=& \Big\{ - \frac12,\, \frac12,\, - \frac12 \Big\}\, , 
\eea
\end{subequations}
where $\theta_W$ is the Weinberg angle.
The $Z'$ boson only has vector couplings to charged fermions,
\be
\sqrt{4 \pi \alpha}\,
\Big\{ C^v_{Z',e},\, C^v_{Z',u},\, C^v_{Z',d} \Big\} 
= \Big\{ -g_z,\, \frac13 g_z,\, \frac13 g_z \Big\}\, .
\ee

For the electron-nucleon DIS process, the main differences between the $B-L$ $Z'$ model and the dark photon model in Ref.~\cite{Hunt-Smith:2023sdz} are  their couplings to the Standard Model fermions. 
While the dark photon has nonzero axial-vector couplings which play an important role in parity-violating electron scattering~\cite{Thomas:2022qhj}, its vector couplings dominate in the nucleon structure functions in Eq.~(\ref{eq:F2-F3}).
The vector couplings of the dark photon and $Z'$ boson to the Standard Model fermions (electron, $u$-type quark, and $d$-type quark) are shown in Table~\ref{tab:DP-Z'} in the case of $\epsilon \to 0$, where $\epsilon$ is the kinetic mixing parameter for the dark photon~\cite{Okun:1982xi}.
While the dark photon couplings are proportional to the fermion electric charges, the $Z'$ boson has universal couplings to all quark flavors~\cite{Appelquist:2002mw}.

\begin{table}[t]
\caption{Couplings of the dark photon and $Z'$ boson to the Standard Model fermions. The second column only shows the vector couplings, as the effect of the axial-vector couplings is negligible.}
%\renewcommand\arraystretch{1.4}
%\begin{ruledtabular}
\begin{tabular}{ccc} \hline
\             &  ~~~~~~~$A_D$ coupling~~~~ &   ~~~~$Z'$ coupling~~~~ \\ \hline 
$u$-quark~~   & $\frac23\, e\, \epsilon$   &   $\frac13 g_z$ \\
$d$-quark~~   & $-\frac13\, e\, \epsilon$  &   $\frac13 g_z$ \\
electron~~    & $- e\, \epsilon$           &   $- g_z$  \\ \hline
\end{tabular}
\label{tab:DP-Z'}
%\end{ruledtabular}
\end{table}

% \newpage
%%%%%%%%%%%%%%%%%%%%%%%%%%%%%%%%%%%%%%%%%%%%%%%%%%%%%%%%%%%%%%%%%%%%
\section{Global QCD analysis}
\label{sec:global-QCD}

For our analysis of the data we use the JAM global QCD analysis framework, which employs Monte Carlo sampling and state-of-the-art uncertainty quantification~\cite{Cocuzza:2021cbi}.
The PDFs are parametrized at the input scale, chosen to be the mass of the charm quark, $Q_0 = m_c$, using the standard form,
\begin{equation}
    f(x,Q_0^2) = N x^\alpha (1-x)^\beta (1 + \gamma\sqrt{x} + \eta x) \, ,
\label{e.param}
\end{equation}
with free shape parameters $\alpha$, $\beta$, $\gamma$, and $\eta$, and normalization $N$, for each parton flavor, $f = q, \bar q$, and $g$ (see Ref.~\cite{Cocuzza:2021cbi} for further details about the parametrization, and Ref.~\cite{Hunt-Smith:2022ugn} for discussion about systematic uncertainties arising from fitting methodologies).
Along with the 36 free parameters this introduces, there are an additional six free parameters corresponding to nuclear off-shell corrections (for electron-deuteron DIS data), for a total of 42 free parameters alongside the $Z'$ parameters $M_{Z'}$ and $g_z$.
The hard scattering kernels and $Q^2$ evolution are evaluated to NLO accuracy.
As in previous JAM analyses, using data resampling we repeatedly fit to data distorted by Gaussian shifts within their quoted uncertainties via $\chi^2$ minimization. 
The resulting 200 replica parameter sets approximate Bayesian samples of the posterior, from which confidence levels for the PDFs and $Z^\prime$ parameters may be estimated.

For a meaningful comparison with the dark photon model, we maximally keep the same settings as those in Ref.~\cite{Hunt-Smith:2023sdz}.
We therefore include an additional contribution to the total $\chi^2$ corresponding to the value of $g-2$ for the muon, although the latest lattice calculations have led to a prediction that differs from the experimental result of $a_{\mu}$ by only $0.9\ \sigma$~\cite{Boccaletti:2024guq}.
While the current value for the anomaly, combining the latest Standard Model  prediction~\cite{Aoyama:2020ynm} and the improved experimental result~\cite{Muong-2:2023cdq} is $\Delta a_{\mu}^{\rm exp} = (249 \pm 48) \times 10^{-11}$, in our analysis we use the same value for the discrepancy between theory and experiment as used in Ref.~\cite{Hunt-Smith:2023sdz}, namely, $\Delta a_{\mu}^{\rm exp} = (251 \pm 59) \times 10^{-11}$~\cite{Muong-2:2021ojo}, in order to make the comparison systematic.

Our baseline study refers to the global fit analysis without the $Z'$ boson, as reported in Ref.~\cite{Hunt-Smith:2023sdz}. 
The inclusive DIS datasets included consist of fixed target proton and deuteron data from SLAC~\cite{Whitlow:1991uw}, BCDMS~\cite{BCDMS:1989qop}, and NMC~\cite{NewMuon:1996fwh, NewMuon:1996uwk}, and collider data from HERA~\cite{H1:2015ubc}. 
For all DIS datasets, we use the cuts $Q^2 > m_c^2$ and $W^2 > 10$~GeV$^2$, as in Ref.~\cite{Cocuzza:2021cbi}.
Other datasets that were added to constrain the PDF parameters include $pp$ and $pd$ Drell-Yan data from the Fermilab NuSea~\cite{NuSea:2001idv} and SeaQuest~\cite{SeaQuest:2021zxb} experiments, and $Z$-boson rapidity data~\cite{CDF:2010vek, D0:2007djv}, $W$-boson asymmetry data~\cite{CDF:2009cjw, D0:2013lql} and jet production data from $p\bar p$ collisions at the Tevatron~\cite{CDF:2007bvv, D0:2011jpq}.
All of the datasets described here are the same as those used in Ref.~\cite{Hunt-Smith:2023sdz}.

In contrast to the dark photon analysis in Ref.~\cite{Hunt-Smith:2023sdz}, no improvement in $\chi^2$ was found with respect to the baseline result when the $Z'$ boson was included in the global fit.
From this results exclusion constraints on the coupling $g_z$ to the $Z'$ boson can be placed at the 95\% confidence level (CL) by imposing an upper bound on $\chi^2$~\cite{ParticleDataGroup:2024cfk},
\be
\chi^2_{Z'}(g_z) - \chi^2_{\rm baseline} = 3.8\, ,
\ee
in the case of one degree of freedom (for a given $M_{Z'}$ value). 
To investigate the effects of potential missing higher order uncertainties, beyond the NLO corrections considered in this analysis, we followed the procedure suggested by the NNPDF collaboration~\cite{NNPDF:2019ubu}.
In particular, we increased the experimental uncertainty on each data point by adding in quadrature the largest variation in the theoretical prediction obtained by varying $Q^2$ for that point by a factor anywhere between $\frac12$ and 2.
Although the total $\chi^2$ necessarily decreased, there was no hint of a preference for the $Z^\prime$ and very little change in the exclusion region --- the largest change was that the exclusion limit went from 0.085 to 0.080 at 160~GeV.

\begin{figure}[t]
\begin{center}
\includegraphics[width=0.8\columnwidth]{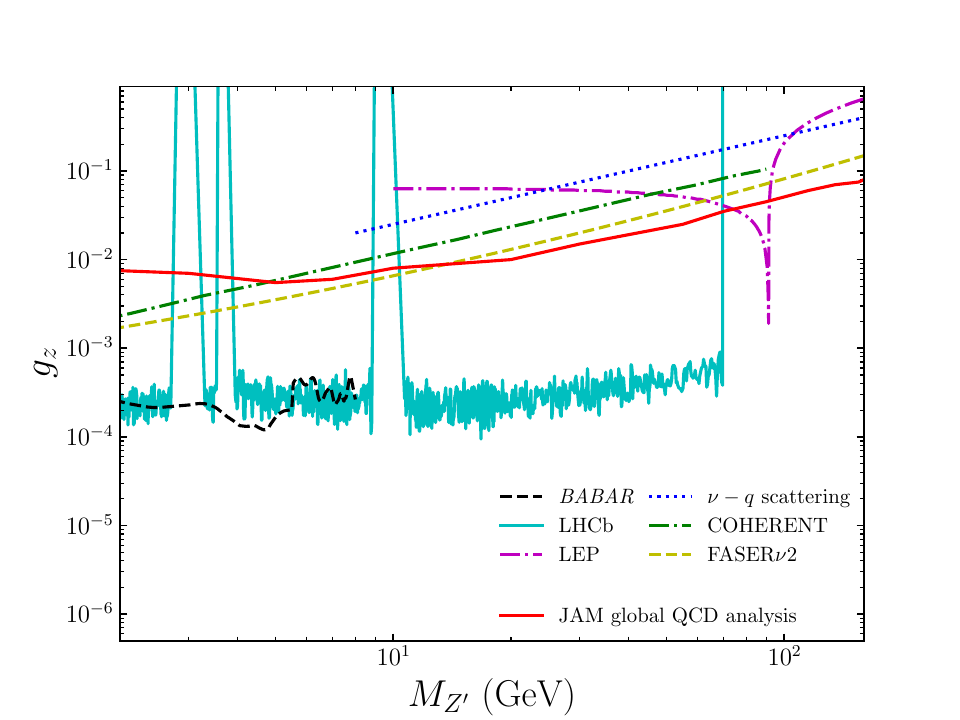}
\caption{Exclusion limits for $g_z$ at the 95\% CL. The {\it BABAR} constraints (90\% CL) are derived by rescaling the limits of the dark photon mixing parameter $\epsilon$~\cite{BaBar:2017tiz}. The LEP and the LHCb limits are taken from Ref.~\cite{Asai:2023mzl}. The upper bound on $g_z$ from neutrino--quark ($\nu-q$) scattering is taken from Ref.~\cite{Heeck:2014zfa}. The exclusion constraints at the 90\% CL from analyses of the COHERENT data and the prospective FASER$\nu$2 data are taken from Refs.~\cite{Cadeddu:2020nbr} and \cite{Asai:2023mzl}, respectively.
}
\label{fig:exlusion}
\end{center}
\end{figure}

In Fig.~\ref{fig:exlusion}, we show the 95\% CL exclusion limits of $g_z$ for a $Z'$ boson with mass in the range %[2, 160]~~GeV. 
between $M_{Z'} = 2$ and 160~GeV.
In this mass region, {\it BABAR} gave strong limits on the mixing parameter $\epsilon$ for the dark photon with mass below 10~GeV~\cite{BaBar:2014zli, BaBar:2017tiz}. 
Those limits can be converted to constraints on the coupling $g_z$ by a rescaling factor $e = \sqrt{4\pi\alpha}$. 
The LHC constraints on the dark photon~\cite{LHCb:2019vmc} can also be converted to those of the $Z'$ boson~\cite{Asai:2023mzl},
\be
g_z^{\rm max}(M_{Z'}\!=\!m_{A'}) = \epsilon^{\rm max}(m_{A'})\, e\, \sqrt{\frac{\sigma(pp \to A')\, {\rm Br}(A' \to \mu^+\mu^-)}{\sigma(pp \to Z')\, {\rm Br}(Z' \to \mu^+\mu^-)}}\, ,
\ee
where $\epsilon^{\rm max}(m_{A'})$ is the upper bound of the kinetic mixing parameter under the assumption that the dark photon only decays to the Standard Model final states.
However, one should bear in mind that, as explained in Sec.~\ref{sec:Intro}, these limits % \sout{, as well as those from the LHC~\cite{LHCb:2019vmc, CMS:2023hwl},} 
could be significantly relaxed in light of the potential couplings of the dark boson to dark matter particles~\cite{Abdullahi:2023tyk}.

Since the $Z$-pole observables are not modified by the $Z'$, the LEP constraints at the 90\% CL only come from the cross sections of the $e^+ e^- \to q \bar{q}$ and $e^+ e^- \to \ell^+ \ell^-$ processes~\cite{Asai:2023mzl}.
These are much weaker than our results in a wide range of the $Z'$ mass.

Neutrino-quark scattering placed limits of order $M_{Z'}/g_z \gtrsim 1$~TeV for $M_{Z'} > 10$~GeV~\cite{Williams:2011qb}, while a more conservative bound of $M_{Z'}/g_z > 0.4$~TeV was adopted in Ref.~\cite{Heeck:2014zfa}, taking into account systematic errors in the NuTeV measurements~\cite{NuTeV:2001whx, Escrihuela:2011cf, Bentz:2009yy}.
Recent analyses of neutrino-nucleus scattering from the COHERENT data~\cite{Cadeddu:2020nbr} and the prospective FASER$\nu$2 data~\cite{Asai:2023mzl} led to improved constraints on $g_z$.
The strongest constraints on the $Z^\prime$ coupling comes from Big Bang nucleosynthesis (BBN), albeit within a specific class of $E_6$ models in which light right-handed neutrinos couple to the $Z'$ boson~\cite{Barger:2003zh}.
Within that framework the limits on $g_z$ are given by $M_{Z'}/g_z > 6.7$~TeV for $M_{Z'} > 10$~GeV and strengthen significantly down to $g_z < 6 \times 10^{-9}$ at $M_{Z'} = 1$~GeV~\cite{Heeck:2014zfa}. 
As Barger {\em et al.}~\cite{Barger:2003zh} observe, there are a number of ways to evade these limits, so we do not show them explicitly in Fig.~\ref{fig:exlusion}.

We also checked the flavored $B_1 - L_1$ model, in which the $Z’$ boson only couples to the first generation of leptons and quarks. 
We repeated the global fit to the high-energy data, with the muon data playing no role in constraining the $Z'$ parameters. 
The resulting exclusion limits on $g_z$ at 95\%~CL become much weaker, with the upper bounds being of ${\cal O}(0.1)$.

%%%%%%%%%%%%%%%%%%%%%%%%%%%%%%%%%%%%%%%%%%%%%%%%%%%%%%%%%%%%%%%%%%%%%%%%%%%%
\section{Conclusion}
\label{sec:conclusion}

We have performed the first global QCD analysis of the world's deep-inelastic and other high-energy scattering data with the inclusion of $U(1)_{B-L}$ gauge boson $Z'$ within the JAM PDF analysis framework~\cite{Hunt-Smith:2023sdz}. 
Our result showed that the inclusion of new gauge bosons does not guarantee an improvement in the $\chi^2$ in a global QCD analysis.
In particular, no improvement in the $\chi^2$ was found with respect to that without the $Z'$, implying that the $B-L$ model is not favored by the data, which included fixed-target and collider DIS, Drell-Yan data, $Z$-boson rapidity and $W$-boson asymmetry data, as well as jet cross sections in hadronic collisions.

Exclusion limits on the coupling to the $Z'$ boson, $g_z$, were then extracted at the 95\%~CL.
These were %{\color{red}\sout{much}} 
stronger than the constraints from neutrino-quark scattering and neutrino-nucleus scattering over the $Z'$ mass range 10 to 160~GeV.
Given that the analysis presented in this work could potentially be extended to other flavor-dependent $U(1)'$ models, such as $B - 3 L_i$, $B_i - L_j$, and their linear combinations~\cite{Chun:2018ibr}, we also explored the flavored $B_1 - L_1$ model, finding a much weaker bound on the corresponding~$g_z$.

Our consideration of the possible impact of missing higher order effects suggested that they were not important for the present analysis.
Nevertheless, it would be worthwhile to include corrections of higher order in $\alpha_s$ in the calculation of the DIS and other scattering cross sections in future work.
On the experimental side, measurements at the proposed Large Hadron electron Collider (LHeC)~\cite{LHeCStudyGroup:2012zhm} and the Future Circular Collider in electron–hadron mode (FCC-eh)~\cite{LHeC:2020van}, as well as the EIC~\cite{AbdulKhalek:2021gbh}, may be expected to place stronger constraints on the $Z'$ parameters~\cite{Kribs:2020vyk}.

%%%%%%%%%%%%%%%%%%%%%%%%%%%%%%%%%%%%%%%%%%%%%%%%%%%%%%%%%%%%%%%%
\begin{acknowledgments}

We would like to thank Arindam Das for helpful communications. This work was supported by the University of Adelaide and the Australian Research Council through the Centre of Excellence for Dark Matter Particle Physics (CE200100008), as well as the DOE contract No.~DE-AC05-06OR23177, under which Jefferson Science Associates, LLC operates Jefferson Lab. The work of N.S. was supported by the DOE, Office of Science, Office of Nuclear Physics in the Early Career Program.

\end{acknowledgments}

% \newpage
%%%%%%%%%%%%%%%%%%%%%%%%%%%%%%%%%%%%%%%%%%%%%%%%%%%%%%%
\bibliography{bibliography}

\end{document}